%% file: main.tex
\documentclass[sigconf]{acmart}

\usepackage{amsfonts,amssymb}

\newcommand{\ignore}[1]{}
\usepackage{xspace}
\newcommand{\paratitle}[1]{\vspace{1.5ex}\noindent\textbf{#1}}
\newcommand{\ie}{\emph{i.e.,}\xspace}

\newcommand{\eg}{\emph{e.g.,}\xspace}

\newcommand{\be}{\mathbf{e}}

\newcommand{\bW}{\mathbf{W}}
\newcommand{\bA}{\mathbf{A}}

\newcommand{\mcal}[1]{\mathcal{#1}}
\newcommand{\mrm}[1]{\mathrm{#1}}
\newcommand{\mbf}[1]{\mathbf{#1}}
\newcommand{\mbb}[1]{\mathbb{#1}}

\usepackage{multirow}
\usepackage{multicol}
\usepackage{caption}
\usepackage{subcaption}
\usepackage{latexsym}
\usepackage{mflogo}
\usepackage{graphics}
\usepackage{tabularx}
\usepackage{balance}

\AtBeginDocument{%
  \providecommand\BibTeX{{%
    \normalfont B\kern-0.5em{\scshape i\kern-0.25em b}\kern-0.8em\TeX}}}

\copyrightyear{2023} 
\acmYear{2023} 
\setcopyright{acmlicensed}\acmConference[SIGIR '23]{Proceedings of the 46th
International ACM SIGIR Conference on Research and Development in
Information Retrieval}{July 23--27, 2023}{Taipei, Taiwan.}
\acmBooktitle{Proceedings of the 46th International ACM SIGIR Conference on
Research and Development in Information Retrieval (SIGIR '23), July 23--27,
2023, Taipei, Taiwan}
\acmPrice{15.00}
\acmDOI{10.1145/3539618.3591786}
\acmISBN{978-1-4503-9408-6/23/07}

\begin{document}




\title{When Search Meets Recommendation: Learning Disentangled Search Representation for Recommendation}

\author{Zihua Si}
\affiliation{%
  \institution{Gaoling School of Artificial
Intelligence\\Renmin University of China}
  \city{Beijing}\country{China}
  }
\email{zihua_si@ruc.edu.cn}

\author{Zhongxiang Sun}
\affiliation{
  \institution{Gaoling School of Artificial
Intelligence\\Renmin University of China}
  \city{Beijing}\country{China}
  }
\email{sunzhongxiang@ruc.edu.cn}

\author{Xiao Zhang}
\authornote{Corresponding authors. Work partially done at Engineering Research Center of Next-Generation Intelligent Search and Recommendation, Ministry of Education.
\\
Work done when Zihua Si and Zhongxiang Sun were interns at Kuaishou.
}
\author{Jun Xu}
\authornotemark[1]
\affiliation{%
  \institution{Gaoling School of Artificial
Intelligence\\Renmin University of China}
  \city{Beijing}\country{China}
  }
\email{{zhangx89, junxu}@ruc.edu.cn}


\author{Xiaoxue Zang}
\author{Yang Song}
\affiliation{%
  \institution{Kuaishou Technology Co., Ltd.}
  \city{Beijing}\country{China}
  }
\email{{zangxiaoxue, yangsong}@kuaishou.com}


\author{Kun Gai}
\affiliation{%
  \institution{Unaffiliated}
  \city{Beijing}\country{China}
  }
\email{gai.kun@qq.com}

\author{Ji-Rong Wen}
\affiliation{%
  \institution{Gaoling School of Artificial
Intelligence\\Renmin University of China}
  \city{Beijing}\country{China}
  }
\email{jrwen@ruc.edu.cn}



\renewcommand{\shortauthors}{Zihua Si et al.}


\begin{abstract}
Modern online service providers such as online shopping platforms often provide both search and recommendation (\textbf{S\&R}) services to meet different user needs. Rarely has there been any effective means of incorporating user behavior data from both S\&R services.  Most existing approaches either simply treat S\&R behaviors separately, or jointly optimize them by aggregating data from both services, ignoring the fact that user intents in S\&R can be distinctively different.
In our paper, we propose a \textbf{S}earch-\textbf{E}nhanced framework for the \textbf{S}equential \textbf{Rec}ommendation (\textbf{SESRec}) that leverages users' search interests for recommendation, by disentangling similar and dissimilar representations within S\&R behaviors.
Specifically, SESRec first aligns query and item embeddings based on users' query-item interactions for the computations of their similarities. 
Two transformer encoders are used to learn the contextual representations of S\&R behaviors independently. 
Then a contrastive learning task is designed to supervise the disentanglement of similar and dissimilar representations from behavior sequences of S\&R.
Finally, we extract user interests by the attention mechanism from three perspectives, \ie the contextual representations, the two separated behaviors containing similar and dissimilar interests. 
Extensive experiments on both industrial and public datasets demonstrate that SESRec consistently outperforms state-of-the-art models.
Empirical studies further validate that SESRec successfully disentangle similar and dissimilar user interests from their S\&R behaviors.
\end{abstract}




\maketitle
\input{introduction}

\input{related}
\input{formulation}
\input{approach}
\input{experiment}
\input{conclusion}

\begin{acks}
This work was funded by the National Key R\&D Program of China (2019YFE0198200),
Beijing Outstanding Young Scientist Program NO. BJJWZYJH012019100020098, Major Innovation \& Planning Interdisciplinary Platform for the ``Double-First Class'' Initiative, and Public Computing Cloud, Renmin University of China.
The work was partially done at Beijing Key Laboratory of Big Data Management and Analysis Methods.
\end{acks}

\bibliographystyle{ACM-Reference-Format}
\balance
\bibliography{main}


\end{document}

%% file: introduction.tex
\section{Introduction}
\label{sec:intro}
Recommender systems and search engines have been widely deployed in online platforms to help users alleviate information overload.
Currently, with the vast increase of data on the Internet, solely using one of the recommender systems or the search engines cannot meet users' information needs.
Hence, many social media platforms, \eg YouTube and TikTok, provide both search and recommendation (\textbf{S\&R}) services for users to obtain information.
As users express their diverse interests in both scenarios, it is feasible to enhance the recommendation system by jointly modeling the behaviors of both, and the core challenge is how to effectively leverage users' search interests for capturing accurate recommendation interests\footnote{In this paper, we use \textbf{recommendation interests} to refer to user interests captured by the recommendation system, and \textbf{search interests} to refer to users' interests revealed in their search history.}.

Early studies~\cite{JSR, JSR2} have demonstrated that jointly optimizing the S\&R models benefits both performances.
Recently, several works~\cite{NRHUB,Query_SeqRec,IV4REC,SRJgraph,USER} have been proposed to boost the recommendation using search data.
As such, devising a search-enhanced framework is a promising research area in the recommendation field. 
Due to that recommendation with search data is still a nascent research area in both academia and industry, recent works~\cite{NRHUB, Query_SeqRec, SRJgraph, USER} usually only incorporate users' S\&R behaviors by feeding them into one encoder to mine users' interests.

\begin{figure}
    \centering
    \begin{subfigure}{0.99\linewidth}
        \centering
        \includegraphics[width=\textwidth]{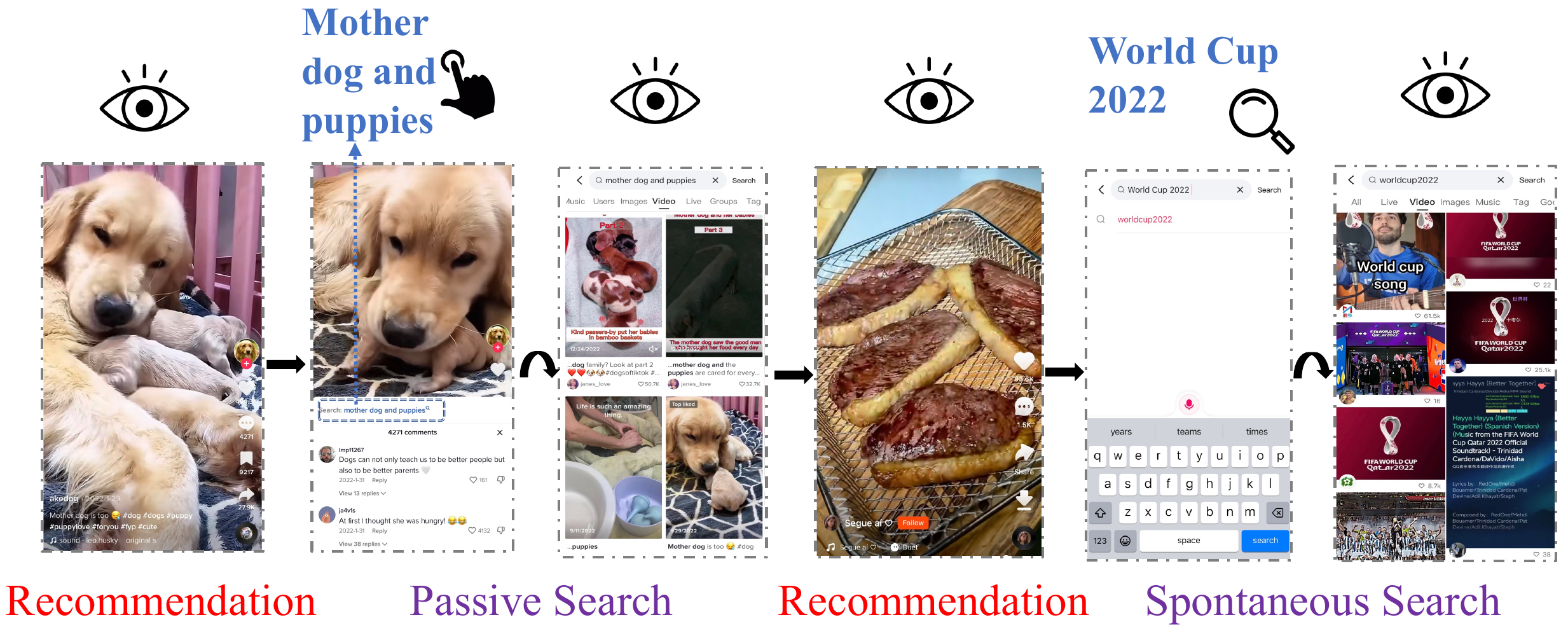}
        \subcaption{An example of user S\&R behaviors.}
    \end{subfigure}
    \begin{subfigure}{0.82\linewidth}
        \centering
        \includegraphics[width=\textwidth]{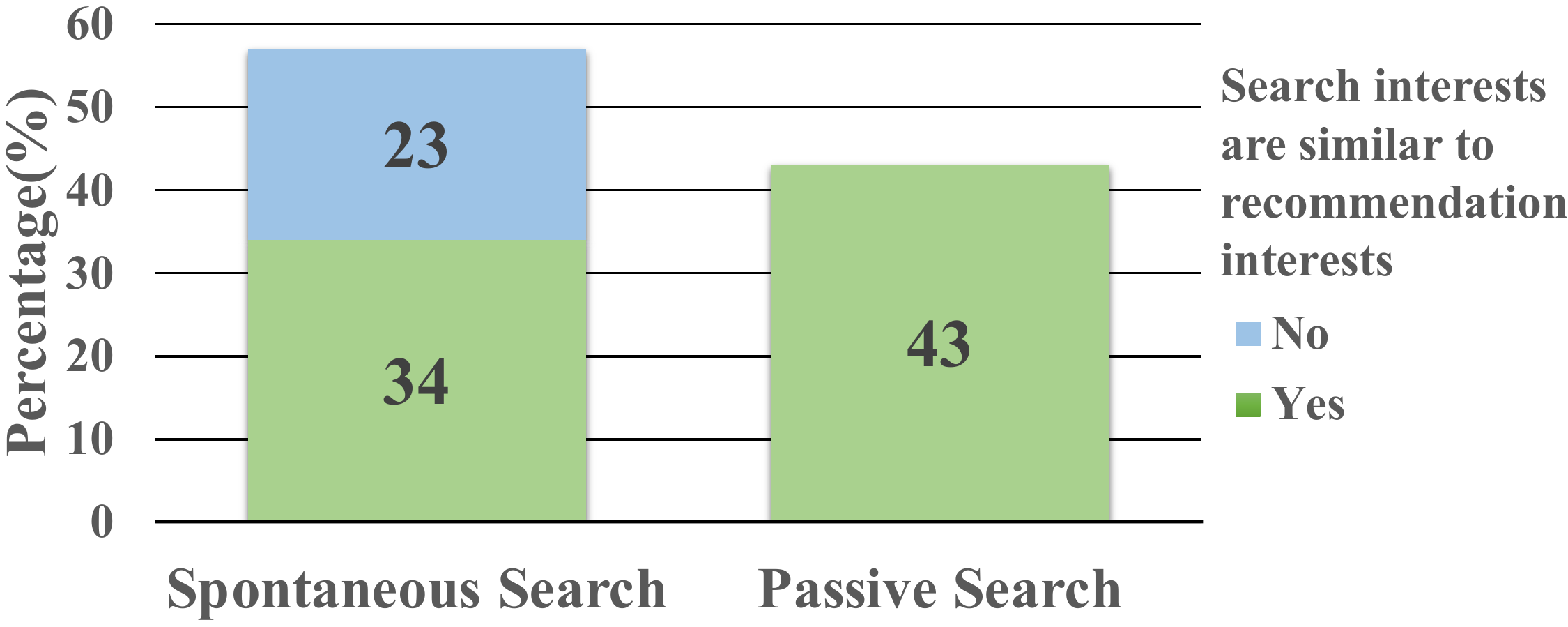}
        \subcaption{Distribution of spontaneous and passive search w.r.t. recommendation interests.}
    \end{subfigure}
    \caption{S\&R behaviors in the short-video scenario.
    (a) 
    After watching a video about dogs, the user chooses to click on the suggested query (passive search) to explore more.
    Later, after watching a food video, the user searches ``world cup 2022'', a spontaneous search unrelated to the watched video.
    (b) Statistics of search behaviors collected from the Kuaishou
    app. 57\% of the search behaviors are spontaneous and 43\% are passive.
    23\% of the spontaneous searches have dissimilar interests to the
    recommendation interests. 
    }
\label{fig:intro}
\end{figure}

Despite their effectiveness, most previous works ignore the differences between users' interests in S\&R behaviors by modeling them without considering their correlations.
However, in real-world applications, search behaviors may strengthen or be complementary to the interests revealed in the recommendation behaviors.
For example, \autoref{fig:intro}(a) illustrates partial behavior histories of a user in the short-video scenario.
While users browsing the items/video suggested by the recommendation system, they may spontaneously start searching by typing queries, which usually differ from the video content in the recommendation feed. We refer to such case as spontaneous search. In contrast, users may also start searching by clicking on the suggested query that is related to the current item/video being played, which we denote as passive search.
To verify the universality of this phenomenon, we conducted data analysis from the real-world data collected from the Kuaishou\footnote{\href{https://www.kuaishou.com/en}{https://www.kuaishou.com/en}} app, shown in~\autoref{fig:intro}(b).
The data analysis is based on behaviors of millions of users.
For each search behavior, if the categories of the items exist in the set of categories of the items interacted by this user in the past seven days, this search behavior is similar to recent recommendation behaviors, and otherwise dissimilar.
The similar search behaviors reflect users' strong interests overlapped in the recommendation behaviors and should be strengthened.
The dissimilar behaviors may be undiscovered interests, which are probably newly emerging and unfulfilled in the recommendation feed.
As a result, it is critical to disentangle the similar and dissimilar representations between S\&R behaviors.

To address such a problem, we devise a search-enhanced framework for sequential recommendation, namely SESRec, to learn the disentangled search representation for the recommendation.
In details, in order to disentangle the similar and dissimilar interests between two behaviors, we propose to decompose each history sequence into two sub-sequences that respectively represent the similar and dissimilar interests so that we can extract user interests from multiple aspects.
To learn the similarity between two behaviors, we first align query-item embeddings with an InfoNCE loss based on users' query-item interactions.
Then two separate encoders are utilized to model S\&R behaviors and generate their contextual representations.
Due to the lack of the labels denoting the similarity of interests between the obtained contextual representations, we propose to leverage self-supervision to guide learning the similar and dissimilar interests. Specifically, 
we exploit the co-attention mechanism to learn the correlation between S\&R's contextual representations. Based on the co-attention score, for both of the contextual representations, we not only aggregate them to generate anchors which are considered to maintain the shared interests of S\&R, but also partition them into two sub-sequences, which are considered to represent the similar and dissimilar behaviors between S\&R (respectively referred to as the positives and the negatives). 
Then following contrastive learning, we define a triplet loss to push the anchors closer to the positives than the negatives.
Finally, we employ the attention mechanism to extract user interests from three aspects, \ie the contextual representations, the positives, and the negatives of S\&R.
In this way, the disentangled interests of S\&R behaviors enhance the prediction for the next interaction.

The contributions of this paper are summarized as follows:

\noindent$\bullet$ To the best of our knowledge, it is the first time that users' S\&R interests are jointly considered and disentangled into similar and dissimilar representations for user modeling.
    We pioneer the practice of learning disentangled search representations for the recommendation.    
    
\noindent$\bullet$ We propose a search-enhanced framework for the sequential recommendation.  
By jointly considering users' S\&R behaviors, we extract users' interests from multiple aspects by decomposing both S\&R behaviors into two parts based on their co-attention scores: one for behaviors containing similar interests and the other for behaviors containing dissimilar interests. 
Moreover, we also utilize self-supervised learning to guide the decomposition.
    
\noindent$\bullet$ We conduct extensive experiments on two datasets. The experimental results validate the effectiveness of our proposed SESRec.
    In particular, SESRec outperforms traditional sequential models which do not leverage search data, as well as other search-aware models, which neglect the correlation between users' interests in S\&R behaviors.
    

%% file: related.tex
\section{RELATED WORK}

\paratitle{Recommendation with Search Data.}
 In both academia and industry, research that enhances recommendation using search data is relatively rare.
Only a few works involve in this area.
\citet{JSR,JSR2} assume that S\&R models could potentially benefit from each other by jointly training with both S\&R data.
\citet{USER} design an approach called USER that mines user interests from the integrated user behavior sequences and accomplishes these two tasks in a unified way.
NRHUB~\cite{NRHUB} exploits heterogeneous user behaviors, \ie webpage browsing and search queries, to enhance the recommendation model.
IV4REC~\cite{IV4REC} and IV4REC+~\cite{IV4Rec+} leverages search queries as instrumental variables to reconstruct user and item embeddings and boost recommendation performance in a causal learning manner.
Query-SeqRec~\cite{Query_SeqRec} is a query-aware model, which incorporates issued queries and browsing items to capture users' interests and intents.
SRJGraph~\cite{SRJgraph} is a GNN-based method which incorporates queries into user-item interaction edges as attributes.
In this work, we also develop a framework to learn disentangled search representation for recommendation.

\paratitle{Sequential Recommendation.}
Sequential recommendation methods mine user interests by modeling sequential relationships of user behaviors.
An early work~\cite{GRU4REC} first utilizes the GRU mechanism to model user preferences.
And attention mechanisms are introduced to capture sequential patterns, such as STAMP~\cite{STAMP}.
There are works using CNN architectures, \eg Caser~\cite{tang2018caser} treats the historical item sequence as an ``image'' and adopts a CNN for user modeling.
For other neural network architectures, several models employ GNN~\cite{chang2021sequential, 10.5555/3367471.3367589} which construct a graph for historical sequences.
Currently, lots of models leverage the transformer architecture, \eg SASRec~\cite{SASREC} and BERT4Rec~\cite{BERT4REC}.
Several works~\cite{DIN,DIEN} devise an attention mechanism to adaptively learn user interests from behaviors.
FMLP-Rec~\cite{FMLPREC} is a state-of-the-art that leverages an all-MLP architecture with learnable filters for sequential recommendation.
Unlike these works, this work incorporates users' search activities into the sequential recommendation task.

\paratitle{Contrastive Learning for Recommendation.}
With the successful development of contrastive learning, this technique has been widely adopted in recommendation~\cite{CORE,S3Rec,10.1145/3394486.3403091,CLSR}.
As for sequential recommendation, \citet{S3Rec} first devise auxiliary self-supervised objectives to enhance data representations via
pre-training methods.
\citet{10.1145/3394486.3403091} propose a sequence-to-sequence training strategy by performing self-supervised learning in the latent space.
Recently, several works~\cite{Re4REC, CLSR} are proposed for learning users' diverse interests.
For example, \citet{Re4REC} propose a contrastive learning framework to disentangle long and short-term interests for recommendation with self-supervision.
In this work, we propose to disentangle the similar and dissimilar interests between S\&R behaviors with self-supervision signals.

%% file: formulation.tex
\section{Problem formulation}
\label{sec: prob form}
Assume that the sets of users, items, and queries are denoted by $\mcal{U}$, $\mcal{I}$, and $\mcal{Q}$ respectively, where $u \in \mcal{U}$ denotes a user, $i \in \mcal{I}$ denotes an item, and $q \in \mcal{Q}$ denotes a query.
For the recommendation history, a user $u$ has a context of chronologically ordered item interactions: $S^u_{i} = [i_1,i_2,\ldots, i_{T_r}]$,
where $S^u_{i}$ is $u$'s interacted item sequence, $T_r$ is the number of items that user $u$ has interacted till timestamp $t$, and $i_k$ is the $k$-th interacted item.
For the search history, a user $u$ has a context of chronologically ordered issued queries: $ S^u_{q} = [q_1, q_2, \ldots, q_{T_s}]$,
where $S^u_{q}$ is $u$'s issued query sequence, $T_s$ is the number of queries issued before $t$, and $q_k$ is the $k$-th issued query.
When using the search service, user $u$ also clicks items after issuing queries: $S^u_{c} = \left[i_{q_1}^{(1)}, i_{q_1}^{(2)}, i_{q_2}^{(1)}, \ldots, i_{T_s}^{(1)}, i_{T_s}^{(2)}, i_{T_s}^{(3)} \right]$,
where $S^u_{c}$ is the $u$'s clicked item sequence corresponding to $S^u_{q}$, and $i_{q_{k}}^{(j)}$ is the $j$-th clicked item under query $q_k$.
The number of the clicked items corresponding to each query can be different and at minimum 0.

Based on the above notations, we define the task of sequential recommendation with search data.
Given the contextual sequences $S^u_{i}$, $S^u_{q}$ and $S^u_{c}$ of a user's recommendation and search histories, the sequential recommendation with search data task aims to predict the next item that the user is likely to interact with at timestamp $t+1$, \ie $ P(i_{t+1} \mid S^u_{i}, S^u_{q}, S^u_{c} )$.
Note that it differs from the conventional sequential recommendation task, which predicts $P(i_{t+1} \mid S^u_{i})$.

%% file: approach.tex
\section{Our Approach: SESRec}
In this section, we elaborate on the proposed SESRec.
\subsection{Overview}

\begin{figure*}
    \centering
        \includegraphics[width=0.90 \linewidth]{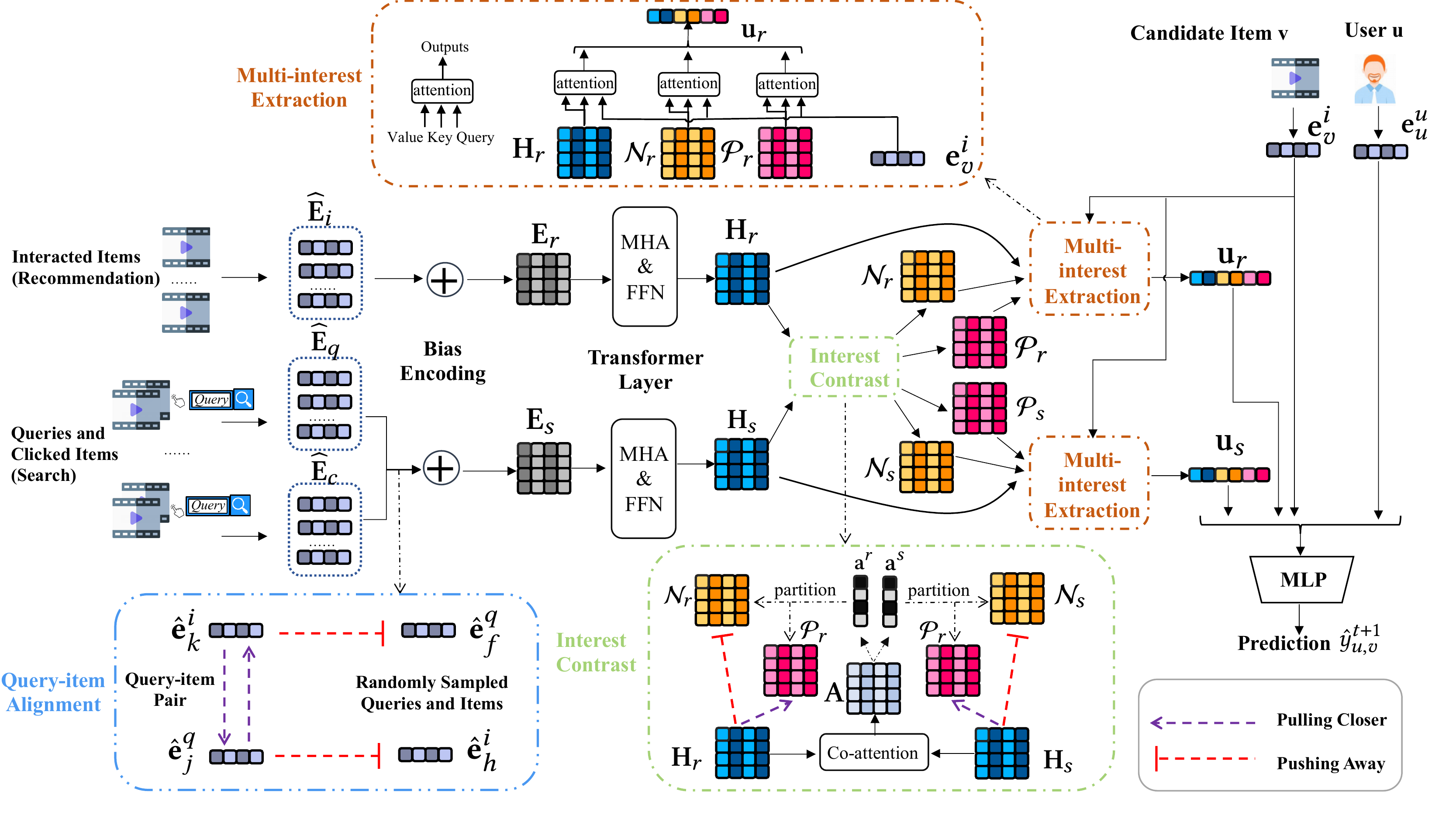}
    \caption{The architecture of SESRec.
    From left to right is the process of modeling S\&R histories.
    On the far right is the process of ultimate prediction.
    The three colored modules with dashed lines conduct interest disentanglement.
    }
\label{fig:model}
\end{figure*}

The overview of SESRec is illustrated in~\autoref{fig:model}.
First, we embed the sparse input data into dense representations.
Then, we leverage the transformer layers~\cite{DBLP:journals/corr/VaswaniSPUJGKP17} to learn the contextual representations of historical behaviors.
For disentangling interests, we separate two behavior sequences into sub-sequences that respectively represent similar and dissimilar interests.  
And we aggregate behavior sequences into vectors to represent user interests w.r.t. the candidate item.
Finally, we concatenate all the vectors together to get the overall representation vector, which is followed by an Multilayer Perceptron (MLP) to generate the ultimate prediction.

Specifically, we design several components to disentangle user interests with self-supervision and aggregate user interests from all aspects.
These designed components are shown in~\autoref{fig:model} within colored boxes.
We align query and item representations into the same semantic space with the InfoNCE loss.
Then we separate S\&R behavior sequences into sub-sequences respectively.
We leverage self-supervision signals to guide the separation based on the triplet loss.
Last, we introduce an interest extraction module that aggregates the original sequences and constructed sub-sequences to form aggregated, similar and dissimilar interest representations of both behaviors.

\subsection{Encoding Sequential Behaviors}
\subsubsection{Embedding Layer}
We maintain separate look-up tables for IDs and attributes of users, items, and queries.
As for users and items, given a user (an item), we concatenate his (its) ID and attribute embeddings to form a user (item) representation:
$\be^u = \be^{\mrm{ID}_u} \|  \be^{a_1} \| \cdots \| \be^{a_n}$ ($\be^i = \be^{\mrm{ID}_i}\| \be^{b_1}\| \cdots \| \be^{b_m}$),
where $\|$ denotes concatenation, $a_1,\ldots,a_n$ and $b_1,\ldots,b_m$ denote the attributes of users and items, respectively.
As for queries, each query $q$ contains several terms ($w_1,w_2,\ldots,w_{|q|}$).
We obtain the query embedding by concatenating query ID embedding and the mean pooling of term embeddings:
$\be^q = \be^{\mrm{ID}_q} \| \mrm{MEAN}(\be^{w_1},\be^{w_2},\ldots,\be^{w_{|q|} })$,
where $\be^{\mrm{ID}_q}$ is the query ID embedding, $\be^{w_{k}}$ is the ID embedding of the $k$-th term, and MEAN$()$ denotes mean pooling.
Many queries occur repeatedly in search data, so query ID is informative.
And most queries consist of less than five terms and lack strong sequential patterns.
So the average pooling operation, following the bag-of-words paradigm, is effective and efficient.

For a user $u$, given the context $S^u_{i}, S^u_{q}$, and $ S^u_{c}$, we obtain embedding matrices of historically interacted items, issued queries, and clicked items, denoted as 
$\mbf E_{i}= [\mbf e^{i}_{1} ,\mbf e^{i}_{2}, \dots,  \mbf e^{i}_{T_r}]^{\intercal} \in \mbb R^{T_r\times d_i}, \mbf E_{q}  = [\mbf e^{q}_{1} ,\mbf e^{q}_{2}, \dots,  \mbf e^{q}_{T_{s}}]^{\intercal} \in \mbb R^{T_s\times d_q}$, and $\mbf E_{c} = [\mbf e^{i}_{1} ,\mbf e^{i}_{2}, \dots,  \mbf e^{i}_{|S_c^u|}]^{\intercal} \in \mbb R^{|S_c^u|\times d_i} $, respectively, where $d_i$ and $d_q$ are dimensions of item and query embeddings.
Besides, we incorporate learnable position encoding matrices to model the sequential order of behaviors, denoted as $\mbf P_s \in \mbb{R}^{T_s\times d}$ and $\mbf P_r \in \mbb{R}^{T_r\times d}$, respectively, where $d$ is dimension.
As for search behaviors, we also adopt type embedding for queries of different sources, \eg user-typed queries, user-specific historical queries, and queries related to the current item.
The search type embedding matrix is denoted by $\mbf M_s \in \mbb{R}^{k\times d}$, where $k$ is the total number of all possible search sources.

Because it is challenging to model user interests with query and item representations in unaligned vector spaces, we transform the item and query embeddings into latent vector spaces with the same dimension.
The transformed embedding matrices of interacted items, issued queries, and clicked items are calculated as:
\begin{equation}
    \widehat{\mbf E}_{i} = \mbf E_{i} \mbf W_{i},\quad
    \widehat{\mbf E}_{q} = \mbf E_{q} \mbf W_{q},\quad
    \widehat{\mbf E}_{c} = \mbf E_{c} \mbf W_{i},
    \label{eq:tranform emb}
\end{equation}
where $\widehat{\mbf E}_{i}  \in \mbb{R}^{T_r\times d}, \widehat{\mbf E}_{q}  \in \mbb{R}^{T_s\times d}$ and $\widehat{\mbf E}_{c}  \in \mbb{R}^{|S^u_c|\times d}$ are transformed matrices, $\mbf W_{i} \in \mbb{R}^{d_i\times d}$ and $\mbf W_{q} \in \mbb{R}^{d_q\times d}$ are trainable parameters for linear projection.

\subsubsection{Bias Encoding}

We incorporate position encodings for S\&R behaviors to make use of the order relations of the sequence.
The recommendation sequence matrix  $\mbf E_r \in \mbb{R}^{T_r\times d}$ is obtained by summing the interacted item matrix and the position matrix:
\begin{equation}
    \mbf E_r = \widehat{\mbf E}_{i} + \mbf P_r
\end{equation}

As for search behaviors, we additionally introduce type encodings along with position encodings.
The search sequence matrix $\mbf E_s \in \mbb{R}^{T_s\times d}$ is defined as:
\begin{equation}
    \mbf E_s = \widehat{\mbf E}_{q} + \widetilde{\mbf E}_{c} + \mbf P_s + \widehat{\mbf M}_s
    \label{eq: search sequence}
\end{equation}
where $\widetilde{\mbf E}_{c} \in \mbb{R}^{T_s\times d}$ is the matrix of the mean pooling of all clicked items' matrices under each query, and $\widehat{\mbf M}_s \in \mbb{R}^{T_s\times d}$ is the type matrix.

To obtain $\widetilde{\mbf E}_{c}$, we first group clicked items $\widehat{\mbf E}_{c}$ by the issued queries, \ie several items clicked by the same query are divided into the same group.
Then we apply a mean pooling operation on each group to get the matrix $\widetilde{\mbf E}_{c} \in \mbb{R}^{T_s\times d}$.

The type matrix $\widehat{\mbf M}_s$ is defined as a sequence of type embeddings for each query in the search history, where each element $\mbf m_j \in \mbb{R}^d$ in $\widehat{\mbf M}_s$ is obtained from the look-up table $\mbf M_s$.
We add the type matrix to model the correlation between search behaviors and search sources.

Considering clicked items contain identical users' interests as their corresponding queries, we fuse the issued query sequence and clicked item sequence to form a unified search sequence by adding them up in Equation~\eqref{eq: search sequence}.

\subsubsection{Transformer Layer}
To learn an enhanced contextual representation for each element in a given sequence, we use the transformer layer~\cite{DBLP:journals/corr/VaswaniSPUJGKP17} to capture the relations between each element with other elements in the S\&R sequences.
The transformer layer generally consists of two sub-layers, \ie a multi-head self-attention layer and a point-wise feed-forward network.
We apply the transformer layers for S\&R sequences, respectively:
\begin{equation}
    \mbf F_s = \mrm{MHA}_s(\mbf E_s),\quad \mbf F_r = \mrm{MHA}_r(\mbf E_r),
\end{equation}
\begin{equation}
    \mbf H_{s} = \mrm{FFN}_s(\mbf F_s),\quad \mbf H_r = \mrm{FFN}_r(\mbf F_r),
\end{equation}
where $\mbf H_{s} \in \mbb{R}^{T_s\times d}$ and $\mbf H_r \in \mbb{R}^{T_r\times d}$ denote enhanced matrices of S\&R sequences respectively, the multi-head self-attention is abbreviated to ``MHA'', and the two-layer feed-forward network is abbreviated to ``FFN''.

\subsection{Self-supervised Interest Disentanglement}
As mentioned before, user interests between S\&R behaviors have overlaps and differences.
Since there does not exist any annotated label of user interests, 
we leverage contrastive learning techniques to disentangle the S\&R behaviors with self-supervision and then extract user interests from three aspects, \ie the aggregated behaviors, the two separated behaviors containing similar and dissimilar interests.

\subsubsection{Query-item Alignment}
It is challenging for the behavior encoders to jointly learn user interests from S\&R behaviors that have unaligned embeddings.
Also, it is unfeasible to disentangle user interests from S\&R behaviors without knowing the semantic similarities between queries and items. Thus, we align the embeddings of queries and items as follows before further extracting user interests from them.


Because items and queries have different forms of features, their original embeddings are unaligned in different vector spaces.
As shown in Equation~\eqref{eq:tranform emb}, we first transform the item and query embeddings into a latent vector space.
Then, inspired by works~\cite{li2021align, radford2021learning} for multi-model learning, we leverage a contrastive learning loss to teach the model which queries and items are similar or different.
Given issued query and clicked item sequence matrices $\widehat{\mbf E}_{q} = [\hat{\mbf e}^q_1,\hat{\mbf e}^q_2,\dots,\hat{\mbf e}^q_{T_s}]^{\intercal} \in \mbb{R}^{T_s\times d}$ and $\widehat{\mbf E}_{c} = [\hat{\mbf e}^i_1,\hat{\mbf e}^i_2,\dots,\hat{\mbf e}^i_{|S^u_c|}]^{\intercal} \in \mbb{R}^{|S^u_c|\times d}$, we minimize the sum of two InfoNCE~\cite{DBLP:journals/corr/abs-1807-03748} losses:
one for query-to-item alignment
\begin{equation}
\mcal{L}_{\mrm{A_{q2i}}}^{u,t} = - \sum_{j=1}^{T_s} \sum_{k=1}^{|q_j|} \log \frac{\exp(s(\hat{\be}^q_j, \hat{\be}^i_k) / \tau )} {\sum_{h \in \mcal{I}_{\mrm{neg}}} \exp( s(\hat{\be}^q_j, \hat{\be}^i_{h}) / \tau)},
\end{equation}
and the other for item-to-query alignment
\begin{equation}
\mcal{L}_{\mrm{A_{i2q}}}^{u,t} = - \sum_{j=1}^{T_s} \sum_{k=1}^{|q_j|} \log \frac{\exp(s(\hat{\be}^q_j, \hat{\be}^i_k) / \tau )} {\sum_{f \in \mcal{Q}_{\mrm{neg}}} \exp( s(\hat{\be}^q_{f}, \hat{\be}^{i}_{k}) / \tau)} ,
\end{equation}
where $\tau$ is a learnable temperature parameter, $|q_j|$ denotes the number of clicked items of query $q_j$ which satisfies $\sum_{j=1}^{T_s}|q_j|=|S^u_c|$, $\mcal{I}_{\mrm{neg}}$ and $\mcal{Q}_{\mrm{neg}}$ denote the sets of randomly sampled items and queries respectively, and $s$ is a similarity function.
The function $s$ is defined as: $s(\mathbf p,\mathbf q) = \tanh( \mathbf p^{\intercal} \bW_{\mrm{A}} \mathbf q)$, where $\tanh$ denotes the activation function and the introduction of $\bW_{\mrm{A}} \in \mbb{R}^{d\times d}$ ensures the query-item correlation estimation can be different with the criterion used in the ultimate prediction.
Finally, the query-item alignment loss is obtained by:
\begin{equation}
    \mcal{L}_{\mrm{ali}}^{u,t} = \frac{1}{2} (\mcal{L}_{\mrm{A_{q2i}}}^{u,t} + \mcal{L}_{\mrm{A_{i2q}}}^{u,t} ),
\end{equation}

\subsubsection{Interest Contrast}
To conduct interest disentanglement, we employ a contrastive learning mechanism to distinguish similar and dissimilar interests between the contextual representations of behaviors $\mbf H_s$ and $\mbf H_r$.

After the transformer layers, given the matrices $\mbf H_s$ and $\mbf H_r$, we construct a co-dependant representation matrix of both behaviors, which generates the similarity scores of two sequences.
Inspired by recent works~\cite{QA1, QA2} for question answering, we leverage the co-attention technique.
We first compute an affinity matrix $\bA \in \mbb{R}^{T_s\times T_r}$ as follows:
\begin{equation}
    \bA = \mrm{tanh}(\mbf H_{s} \bW_{l} (\mbf H_{r})^{\mrm{T}}),
\end{equation}
where $\bW_{l} \in \mbb{R}^{d\times d}$ is a learnable weight matrix.
The affinity matrix $\mbf A$ contains affinity scores corresponding to all pairs of recommendation behaviors and search behaviors.
We multiply the affinity matrix $\bA$ and the search matrix $\mbf H_s$ (or the recommendation matrix $\mbf H_r$), and then normalize the multiplication results to get similarity scores for each element in one sequence across all the elements in the other sequence:
\begin{equation}
    \mbf a^{s} = \mrm{softmax}( \bW_{r}\mbf H_{r}^{\mrm T} \bA^{\mrm T} ),\quad \mbf a^{r} = \mrm{softmax}(  \bW_{s}\mbf H_{s}^{\mrm T}\bA ),
\end{equation}
where $\mbf a^{r} \in \mbb{R}^{T_r}$ and $\mbf a^{s} \in \mbb{R}^{T_s}$ are similarity scores, $\bW_{r},\bW_{s} \in \mbb{R}^{1\times d}$ are trainable parameters for linear projection.

Next, we exploit a triplet loss to self-supervise the disentanglement of similar and dissimilar interests between two behaviors.
Given similarity scores $\mbf a^{r}$ and $\mbf a^{s}$, elements in $\mbf H_s$ and $\mbf H_r$ with higher scores can be interpreted as representative ones for similar interests, while elements with lower scores can be interpreted as representative ones for dissimilar interests.
Let $\mcal{P}$ ($\mcal{N}$ ) denote the set of elements containing similar (dissimilar) interests of S\&R behaviors.
As such, we perform hard selection to separate S\&R sequences into two subsequences as follows:
\begin{equation}
    \mcal{P}_s = \{\mbf h^s_{j} \mid \mbf a^s_{j} > \gamma_s \}, \quad \mcal{N}_s = \{\mbf h^s_{j} \mid \mbf a^s_{j} \leq \gamma_s\}, 
    \label{eq:separated1}
\end{equation}
\begin{equation}
    \mcal{P}_r = \{\mbf h^r_{j} \mid \mbf a^r_{j} > \gamma_r\}, \quad \mcal{N}_r = \{\mbf h^r_{j} \mid \mbf a^r_{j} \leq \gamma_r\},
    \label{eq:separated2}
\end{equation}
where $\mbf h^s_{j},\mbf h^r_{j} \in \mbb{R}^{d}$ are the $j$-th vectors in matrices $\mbf H_s$ and $\mbf H_r$, $\mbf a^s_{j}$ and $\mbf a^r_{j}$ are similarity scores for $\mbf h^s_{j}$ and $\mbf h^r_{j}$ respectively, $\gamma_r$ and $\gamma_s$ are selection thresholds.
Since $\mbf a^s$ and $\mbf a^r$ are normalized after $\mrm{softmax}$, we empirically set the thresholds $\gamma_r$ and $\gamma_s$ to the uniform values $\frac{1}{T_r}$ and $\frac{1}{T_s}$.
The positives with similarity scores larger than the thresholds can be interpreted as similar interests with above-average similarities.
The negatives, as the counterparts of positives, are with below-average similarities.

Then we design the anchors, positives and negatives of the triplet loss.
To guide learning the disentanglement, we utilize the original sequences $\mbf H_r$ and $\mbf H_s$ to form anchors, and leverage the separated subsequences $\mcal{P}_s, \mcal{P}_r$ and $\mcal{N}_s, \mcal{N}_r$ to serve as positives and negatives.
The anchors, positives, and negatives can be calculated as:
\begin{equation}
    \mbf i_s^A=\sum_{j=1}^{T_s} \mbf a_j^s \mbf h^s_{j},\quad \mbf i_s^P = \mrm{MEAN}(\mcal{P}_s), \quad \mbf i_s^N = \mrm{MEAN}(\mcal{N}_s),
\end{equation}
\begin{equation}
    \mbf i_r^A=\sum_{j=1}^{T_r} \mbf a_j^r \mbf h^r_{j},\quad \mbf i_r^P = \mrm{MEAN}(\mcal{P}_r), \quad \mbf i_r^N = \mrm{MEAN}(\mcal{N}_r),
\end{equation}
where $\mbf i_s^A, i_r^A \in \mbb{R}^{d}$ are anchors, $\mbf i_s^P, \mbf i_r^P \in \mbb{R}^{d}$ are positives, $\mbf i_s^N,\mbf i_r^N \in \mbb{R}^{d}$ are negatives.
Then we perform contrastive learning, which requires the anchors to be similar with positives, and to be different from negatives.
Based on these vectors, We implement triplet losses for S\&R behaviors, respectively.
Formally, the loss function is computed as follows:
\begin{equation}
    \mcal{L}_{\mrm{tri}}(a,p,n) = \max \{d(a,p)-d(a,n)+m, 0\},
\end{equation}
where $d$ denotes distance function which is implemented as euclidean distance, $m$ denotes a positive margin value, $a,p$ and $n$ denote anchors, positives and negatives, respectively.
Finally, the interest contrast loss can be obtained by summing up two triplet losses,
one for recommendation behaviors, the other for search behaviors:
\begin{equation}
\mcal{L}_{\mrm{con}}^{u,t} =  \mcal{L}_{\mrm{tri}}(\mbf i_r^A, \mbf i_r^P, \mbf i_r^N) + \mcal{L}_{\mrm{tri}}(\mbf i_s^A, \mbf i_s^P, \mbf i_s^N),
\end{equation}


\paratitle{Remark.}
In most cases, users use S\&R services at different frequencies.
The lengths and update frequencies of two behaviors are different since they are collected from different services.
That is why we employ triplet losses for the two behaviors, respectively.
We update the model parameters of each behavior with its own constructed interest representations, which ensures the consistency of model training.
Besides, considering that similar and dissimilar interests usually overlap with each other to some extent, there is no clear distinction between them.
The triplet loss performs pairwise comparisons, which reduces the differences between similar things and increases the differences between different things.
That is why we use the triplet loss instead of other contrastive loss functions, \eg InfoNCE~\cite{DBLP:journals/corr/abs-1807-03748}, which imposes too strong punishment on the similarity between positives and negatives.

\subsubsection{Multi-interest Extraction}
Based on the original behaviors and separated behaviors containing similar and dissimilar interests, we extract user interests from three aspects, \ie aggregated, similar, and dissimilar interests.
Given a candidate item $v$, we utilize an attention mechanism to reallocate the user interests w.r.t. the candidate item.
For recommendation behaviors, interests can be extracted from three aspects as follows:
\begin{equation}
    \mbf u_{r}^{\mrm{all}} = \sum_{j=1}^{T_r} a_j^{\mrm{all}} \mbf h_j^r,
    \quad a_j^{\mrm{all}} = \frac{\exp( (\mbf h_j^r)^{\mrm{T}} \mbf W_{d} \mbf e_v^i)}{
    \sum_{k=1}^{T_r} \exp((\mbf h_k^r)^{\mrm{T}} \mbf W_{d} \mbf e_v^i)},
\end{equation}

\begin{equation}
    \mbf u_{r}^{\mrm{sim}} = \sum_{\mbf h_j^r\in \mcal{P}_r} a_j^{\mrm{sim}} \mbf h_j^r,
    \quad a_j^{\mrm{sim}} = \frac{\exp( (\mbf h_j^r)^{\mrm{T}} \mbf W_{d} \mbf e_v^i)}{
    \sum_{\mbf h_k^r \in \mcal{P}_r} \exp((\mbf h_k^r)^{\mrm{T}} \mbf W_{d} \mbf e_v^i)},
\end{equation}

\begin{equation}
    \mbf u_{r}^{\mrm{diff}} = \sum_{\mbf h_j^r \in \mcal{N}_r} a_j^{\mrm{diff}} \mbf h_j^r,
    \quad a_j^{\mrm{diff}} = \frac{\exp( (\mbf h_j^r)^{\mrm{T}} \mbf W_{d} \mbf e_v^i)}{
    \sum_{\mbf h_k^r \in \mcal{N}_r} \exp((\mbf h_k^r)^{\mrm{T}} \mbf W_{d} \mbf e_v^i)},
\end{equation}
where $\mbf u_{r}^{\mrm{all}}, \mbf u_{r}^{\mrm{sim}}, \mbf u_{r}^{\mrm{diff}} \in \mbb{R}^{d}$ are representative vectors for aggregated, similar, and dissimilar interests, $\mbf W_d$ is the trainable parameters to model correlation between recommendation behaviors and the candidate item.
By concatenating these three vectors, we can get the representation of recommendation interests:
\begin{equation}
    \mbf u_r = \mbf u_{r}^{\mrm{all}} \|\mbf u_{r}^{\mrm{sim}} \| \mbf u_{r}^{\mrm{diff}},
\end{equation}
where $\mbf u_r \in \mbb{R}^{3d}$.
Similarly, we can obtain the representation of search interests in the same way, \ie $\mbf u_s \in \mbb{R}^{3d}$.

\subsection{Prediction and Model Training}
\subsubsection{Prediction.}
To predict the interaction, we utilize the widely adopted two-layer MLP~\cite{DIN, DIEN} to model feature interaction and make predictions.
Given a user $u$ and an item $v$ at timestamp $t+1$, the prediction score can be calculated as follows:
\begin{equation}
    \hat{y}^{t+1}_{u,v} = \mrm{MLP}(\mbf u_{r} \| \mbf u_s \| \mbf e_v^i \| \mbf e_u^u),
\end{equation}
where $\hat{y}^{t+1}_{u,v}$ denotes the prediction score, $\mbf e_v^i$ and $\mbf e_u^u$ are embeddings of the item $v$ and the user $u$, respectively. 

\subsubsection{Model Training.}
Following the existing works' settings~\cite{DIN, DIEN}, we adopt the negative log-likelihood function to supervise the final prediction:
\begin{equation}
    \mcal{L}^{u,t}_{\mrm{rec}} = -\frac{1}{N}\sum_{v\in \mcal{O}} y_{u,v}^{t+1}\log(\hat{y}^{t+1}_{u,v})+(1-y^{t+1}_{u,v})\log(1-\hat{y}^{t+1}_{u,v}),
\end{equation}
where $\mcal{O}$ is the set composed of training pairs of one positive item and $N-1$ negative items.
In order to apply additional self-supervised signals about query-item alignment and interest disentanglement, we train our model in an end-to-end manner under a multi-task learning schema.
The overall loss function is formulated as:
\begin{equation}
    \mcal{L} = \sum_{u=1}^{|\mcal{U}|}\sum_{t=1}^{T_u}(\mcal{L}^{u,t}_{\mrm{rec}} + \alpha \mcal{L}_{\mrm{ali}}^{u,t} + \beta \mcal{L}_{\mrm{con}}^{u,t}) + \lambda|| \Theta ||_2.
    \label{eq: loss func}
\end{equation}
where $|\mcal{U}|$ is the number of users, $T_u$ denotes the timestamp of the user $u$'s latest interaction,  $\alpha$ and $\beta$ are hyper-parameters for additional tasks, and $\lambda || \Theta ||_2$ denotes the $L_2$ regularization to avoid over-fitting. 

%% file: experiment.tex
\section{Experiment}
\subsection{Experimental Setup}

\subsubsection{Dataset.}

SESRec needs user S\&R behavior logs simultaneously.
In the following experiments, we evaluated models on two datasets: one is collected from logs of a short-video app, and the other is based on a widely used public Amazon dataset~\cite{amazon_dataset, amazon_dataset2}.
\autoref{tab: dataset} reports statistics of both datasets.

\begin{table}[t]
 \caption {Statistics of datasets used in this paper. `S' and `R' denote search and recommendation, respectively.}\label{tab: dataset} 
 \vspace{-10px}
{
\tabcolsep=0.09cm 
\begin{tabular}{llllll}
\hline
 Dataset & Users & Items & Queries & Actions-S & Actions-R \\ \hline
 Kuaishou & 35,721  & 822,832 &398,924 &922,531 &11,381,172   \\
 Amazon & 68,223  & 61,934 & 4,298 & 934,664 & 989,618   \\ \hline
\end{tabular}
}
\vspace{-12px}
\end{table}

\begin{table*}[h!]
    \caption{Overall performance comparisons on both datasets.
    The best and the second-best performance methods are denoted in bold and underlined fonts respectively.
    * means improvements over the second-best methods are significant (\textit{p}-value < $0.01$).}
    \vspace{-5px}
    \label{exp:overall}
    \resizebox{0.99\textwidth}{!}{
         \begin{tabular}{cc|cccccc|cccccc}
          \toprule
          \multicolumn{2}{c|}{Dataset} & \multicolumn{6}{c|}{Kuaishou} & \multicolumn{6}{c}{Amazon (Kindle Store)} \\
          \hline
          Category & Method & NDCG@5 & NDCG@10 & HIT@1 & HIT@5 & HIT@10 & MRR & NDCG@5 & NDCG@10 & HIT@1 & HIT@5 &HIT@10 &MRR\\
          \midrule
          \multicolumn{1}{c}{\multirow {6}{*}{\shortstack{Sequential}} } 
          & STAMP & 0.2544 & 0.2981 & 0.1413 & 0.3616 & 0.4970 & 0.2569 
          & 0.2612 & 0.3103 & 0.1336 & 0.3833 & 0.5352 & 0.2608 \\
          & DIN & 0.2969 & 0.3418 & 0.1792 & 0.4092 & 0.5484 & 0.2976 
          &0.2999  &0.3495 &0.1591 &0.4340 &0.5871 & 0.2942   \\
          & GRU4Rec & 0.3247 & 0.3688 & 0.1890 & 0.4517 & 0.5881 & 0.3180 
          & 0.3099 & 0.3662 & 0.1479 & 0.4648 & 0.6388 & 0.2993 \\
          & SASRec & 0.3252 & 0.3693 & 0.1904 & 0.4501 & 0.5864 & 0.3187  
          & 0.3822 & 0.4312 & 0.2187 & 0.5324 &0.6838 & 0.3675\\
          & DIEN & 0.3217 & 0.3704 & 0.1914 & 0.4463 &0.5969  & 0.3192 
          & 0.3336 & 0.3803 & 0.1871 & 0.4706 & 0.6150 & 0.3246  \\
          & FMLP-Rec & \underline{0.3354} & \underline{0.3787} & 0.1953 & \underline{0.4651} & \underline{0.5988} & 0.3270
          & \underline{0.4073} & \underline{0.4550} & 0.2349 & \underline{0.5651} & \textbf{0.7121} & \underline{0.3883} \\
          \hline
          \multicolumn{1}{c}{\multirow {6}{*}{\shortstack{Search-aware} } } 
          & NRHUB & 0.2964 & 0.3431 & 0.1665 & 0.4199 & 0.5647 & 0.2933 
          & 0.2744 & 0.3265 & 0.1329 & 0.4099 & 0.5708 & 0.2704  \\
          & JSR & 0.3015 &0.3513  & 0.1738 & 0.4241 & 0.5783 &  0.3004
          & 0.3221 & 0.3722 & 0.2057 & 0.4386 & 0.5937 & 0.3224   \\
          & IV4REC & 0.3114 & 0.3591 & 0.1877 & 0.4282 & 0.5761 & 0.3116 & 0.3473 & 0.3960 & 0.1853 & 0.4985 & 0.6258 & 0.3331  \\
          & Query-SeqRec & 0.3117 & 0.3581 & 0.1740 &0.4412  & 0.5844 & 0.3055 & 0.3692 & 0.4142 & 0.2187 & 0.5083 & 0.6470 & 0.3572 \\
          & SRJGraph & 0.3297 & 0.3762 & \underline{0.2046} & 0.4479 & 0.5917 & \underline{0.3277} & 0.3670 & 0.4043 & \textbf{0.2760} &0.4898 &0.6242 &0.3708 \\
          & SESRec & \textbf{0.3541}* & \textbf{0.4054}* & \textbf{0.2173}* & \textbf{0.4848}* & \textbf{0.6436}* & \textbf{0.3490}* & \textbf{0.4224}*  & \textbf{0.4663}* & \underline{0.2580} & \textbf{0.5723}* & \underline{0.7074} & \textbf{0.4046}*\\
          \bottomrule
        \end{tabular}
    }
    \vspace{-5px}
\end{table*}

\textbf{Kuaishou Dataset}:
This dataset is constructed based on behavior logs of 35,721 users who elected to use both S\&R services on the short-video app named Kuaishou over one month in 2022.
The historical S\&R behaviors have been recorded.
For dataset pre-processing, following the common practice in~\cite{FMLPREC, BERT4REC, SASREC},
we group interaction records by users, sort them by timestamp ascendingly and filter unpopular items and users with fewer than five interaction records.

\textbf{Amazon Dataset}\footnote{The Amazon review dataset can be found at \href{http://jmcauley.ucsd.edu/data/amazon/}{http://jmcauley.ucsd.edu/data/amazon/}.}:
To the best of our knowledge, there doesn't exist a public dataset that contains both S\&R behaviors.
Following a standard approach for product search~\cite{amazon_search_process}, we enhance a recommendation dataset, the Amazon dataset~\cite{amazon_dataset, amazon_dataset2}, by generating search behaviors.
We adopt the ``Kindle Store'' subset of the five-core Amazon dataset that covers data in which all users and items have at least five reviews.
The detailed generation process of search behaviors\footnote{The constructed search data is available at \href{https://github.com/QingyaoAi/Amazon-Product-Search-Datasets}{https://github.com/QingyaoAi/Amazon-Product-Search-Datasets}.} can be found in~\cite{amazon_search_process}.
Please note that this automatically constructed search dataset has been widely used by product search community~\cite{amazon_search_process, 10.1145/3295822, 10.1145/2983323.2983702, JSR2}.
Following~\cite{10.1145/3295822}, we randomly select one query for items with multiple constructed queries to model the sequential behaviors.

Following previous works~\cite{FMLPREC, BERT4REC, SASREC}, we adopt the \textit{leave-one-out} strategy to split both datasets.
For each user, we hold out the most recent action for testing, the second most recent action for validation, and all the remaining actions for training.
Besides, to ensure data quality, we filter interactions in which the user doesn't have historical S\&R history simultaneously.

\subsubsection{Evaluation Metrics.}

Following~\cite{BERT4REC, FMLPREC}, we employ several widely used ranking metrics, including \textit{Hit Ratio} (HIT), \textit{Normalized Discounted Cumulative Gain} (NDCG), and \textit{Mean Reciprocal Rank} (MRR).
We report HIT with $k=1,5,10$ and NDCG with $k=5,10$.
We pair the ground-truth item with 99 randomly sampled items that the user has never interacted with.
For all metrics, we calculate them according to the ranking of items and report average results.

\subsubsection{Baseline Models.}

In this work, we compare SESRec with state-of-the-art methods.

For sequential recommendation methods without leveraging search data, we include following \textbf{\emph{sequential}} models:
(1) \textbf{STAMP}~\cite{STAMP}: It captures users' general interests from the long-term memory and short-term memory;
(2) \textbf{DIN}~\cite{DIN}: It uses an attention mechanism to model user interest from historical behaviors w.r.t. a target item;
(3) \textbf{GRU4Rec}~\cite{GRU4REC}: It is the first work to apply RNN to session-based recommendation with a ranking based loss;
(4) \textbf{SASRec}~\cite{SASREC}: It is a unidirectional transformer-based sequential model, which uses self-attention to capture sequential preferences;
(5) \textbf{DIEN}~\cite{DIEN}: It enhances DIN by combining attention with GRU units to take interests evolution into consideration;
(6) \textbf{FMLPRec}~\cite{FMLPREC}: It is an all-MLP model with learnable filters which can adaptively attenuate the noise information in historical sequences.
 
For methods using search data, we include following \textbf{\emph{search-aware}} models:
(7) \textbf{NRHUB}~\cite{NRHUB}: It is a news recommendation model leveraging heterogeneous user behaviors;
(8) \textbf{JSR}~\cite{JSR}: It is a general framework which optimizes a joint loss.
We implement it following~\cite{IV4REC} to ensure a fair comparison with other sequential models;
(9) \textbf{IV4REC}~\cite{IV4REC}: It is a model-agnostic framework exploiting search queries as instrumental variables to enhance the recommendation model.
Following the original paper, we apply this framework over DIN;
(10) \textbf{Query-SeqRec}~\cite{Query_SeqRec}: It is a query-aware sequential model which incorporates queries into user behaviors using a transformer-based model.
(11) \textbf{SRJGraph}~\cite{SRJgraph}: It is a GNN-based model which exploits a heterogeneous graph to model the user-item and user-query-item interactions for S\&R.

\subsubsection{Implementation Details.}

All hyper-parameters of baselines are searched following suggestions from the original papers.
For all models, the maximum sequence length of recommendation (search) history is set to 150 (25) on the Kuaishou dataset and 15 (15) on the Amazon dataset.
$d_i,d_q$ and $d$ are set as 48, 64 and 48 (32, 32, and 32) on the Kuaishou dataset (Amazon dataset).
For the fair competition, we deploy the same setting of item embeddings on all models.
For query embeddings, we also randomly initialize term embeddings for all search-aware models.
The batch size is set as 256.
The hyper-parameters $\alpha$ and $\beta$ are set as $0.1$ and $0.001$, respectively.
The margin $m$ is set as $0.1$.
We use the Adam~\cite{DBLP:journals/corr/KingmaB14} with a learning rate of 0.001, and adopt early-stopped training to avoid over-fitting.
More details can be found in the open source codes\footnote{ \href{https://github.com/Ethan00Si/SESREC-SIGIR-2023}{https://github.com/Ethan00Si/SESREC-SIGIR-2023}}.

\subsection{Overall Performance}

\autoref{exp:overall} reports the recommendation results on the two datasets.
We have the following observations:

\noindent$\bullet$ Search-aware models do not always bring performance gains. 
SRJGraph is the SOTA approach that mines both S\&R behaviors.
However, the SOTA sequential model FMLP-Rec can obtain compatible or even better performance than SRJGraph.
Besides, Query-SeqRec shares a similar architecture as SASRec but achieves slightly poorer performance than SASRec.
These phenomenons indicate that blindly incorporating S\&R histories may be insufficient to capture users' diverse interests because users' interests in essence are entangled. 

\noindent$\bullet$ Compared to baseline models, SESRec achieves the best performance on both datasets in most cases, significantly outperforming the SOTA sequential and search-aware methods FMLP-Rec and SRJGraph by a large margin (paired t-test at $p$-value $<0.01$).
The relative improvements over conventional sequential models reveal that leveraging search behaviors can boost recommendation models.
Furthermore, the substantial performance gains over search-aware models validate the effectiveness of interest disentanglement in S\&R behaviors.

\noindent$\bullet$ Comparing SESRec on two datasets, SESRec achieves fewer relative improvements on the Amazon dataset.
Considering that search data of the Amazon dataset was automatically constructed, many items share the same queries, and the number of queries is sparse compared with the number of items, as shown in~\autoref{tab: dataset}.
Thus, the query-item alignment and interest contrast modules play a minor role in boosting recommendation performance on this dataset.


\subsection{Detailed Empirical Analysis}
In this section, we conducted more detailed experiments on the real-world Kuaishou dataset, providing in-depth analyses of how and why SESRec achieved state-of-the-art performance. 

\subsubsection{Ablation Study.}
\label{sec:ablation}
SESRec consists of several key components, including alignment for queries and items, disentanglement for user S\&R interests with self-supervised signals, and the multi-interest extraction module.
To investigate how different components affect the performance of SESRec, we conducted ablation studies by progressively adding three components to the base model.
We added these modules one by one because each module depends on previous modules.
The base model solely processes S\&R behaviors with transformer layers and the interest extraction module of aggregated interests.
\autoref{tab:ablation} shows the results on the Kuaishou dataset.
Next, we give a detailed discussion about each component:

\begin{table}
\centering
    \caption{Ablation studies by progressively adding proposed modules to the base model. MIE is short for the multi-interest extraction module.}
    \vspace{-10px}
{
\renewcommand{\arraystretch}{1.1}
\begin{tabular}{l ccc ccc}
\toprule
Model
   & N@5 & N@10 & H@1 & H@5 &H@10 &MRR \\
    \midrule
    Base    &   0.3394   &   0.3770   &   0.2027  &  0.4618   &  0.6094    &   0.3294       \\
    $+\mcal{L}_{\mrm{ali}}^{u,t}$   &    0.3464  &  0.3982   &  0.2106   &   0.4762  &  0.6308  &   0.3421       \\
    \ \ $+\mcal{L}_{\mrm{con}}^{u,t}$    &   0.3507   &   0.4021   &  0.2139     &   0.4812   &  0.6406    &   0.3459        \\
    $\quad+\mrm{MIE}$    &   \textbf{0.3541}   &   \textbf{0.4054}   &   \textbf{0.2173}    &   \textbf{0.4848}   &  \textbf{0.6436}    &  \textbf{0.3490}         \\
    \bottomrule
\end{tabular}
}%
    \label{tab:ablation}
\end{table}

\noindent$\bullet$ $\mcal{L}_{\mrm{ali}}^{u,t}$: 
denotes the loss function of query-item alignment, which guarantees that model captures the correlation between queries and items.
We observed that adding $\mcal{L}_{\mrm{ali}}^{u,t}$ leads to consistent performance gain.
The results demonstrate that understanding the interactions between queries and items is beneficial to jointly model S\&R behaviors.

\noindent$\bullet$ $\mcal{L}_{\mrm{con}}^{u,t}$:
refers to the loss function of interest contrast, which is designed to disentangle similar and dissimilar interests between S\&R behaviors.
The interest disentanglement leads to performance improvement, which indicates the necessity of disentanglement.
We attribute the improvement to the fact that $\mcal{L}_{\mrm{con}}^{u,t}$ helps the model capture more accurate representations of user interests.

\noindent$\bullet$ MIE: is short for the multi-interest extraction module, which is designed to extract interests from three perspectives, \ie aggregated, similar, and dissimilar interests.
We found that MIE contributes to the final prediction, validating the importance of MIE.
Though $\mcal{L}_{\mrm{con}}^{u,t}$ has distinguished interests into similar and dissimilar ones, the model cannot explicitly leverage similar or dissimilar interests for prediction without MIE.

\subsubsection{Analysis of Disentangled Interests.}

\begin{figure}
    \centering
        \includegraphics[width=0.9\linewidth]{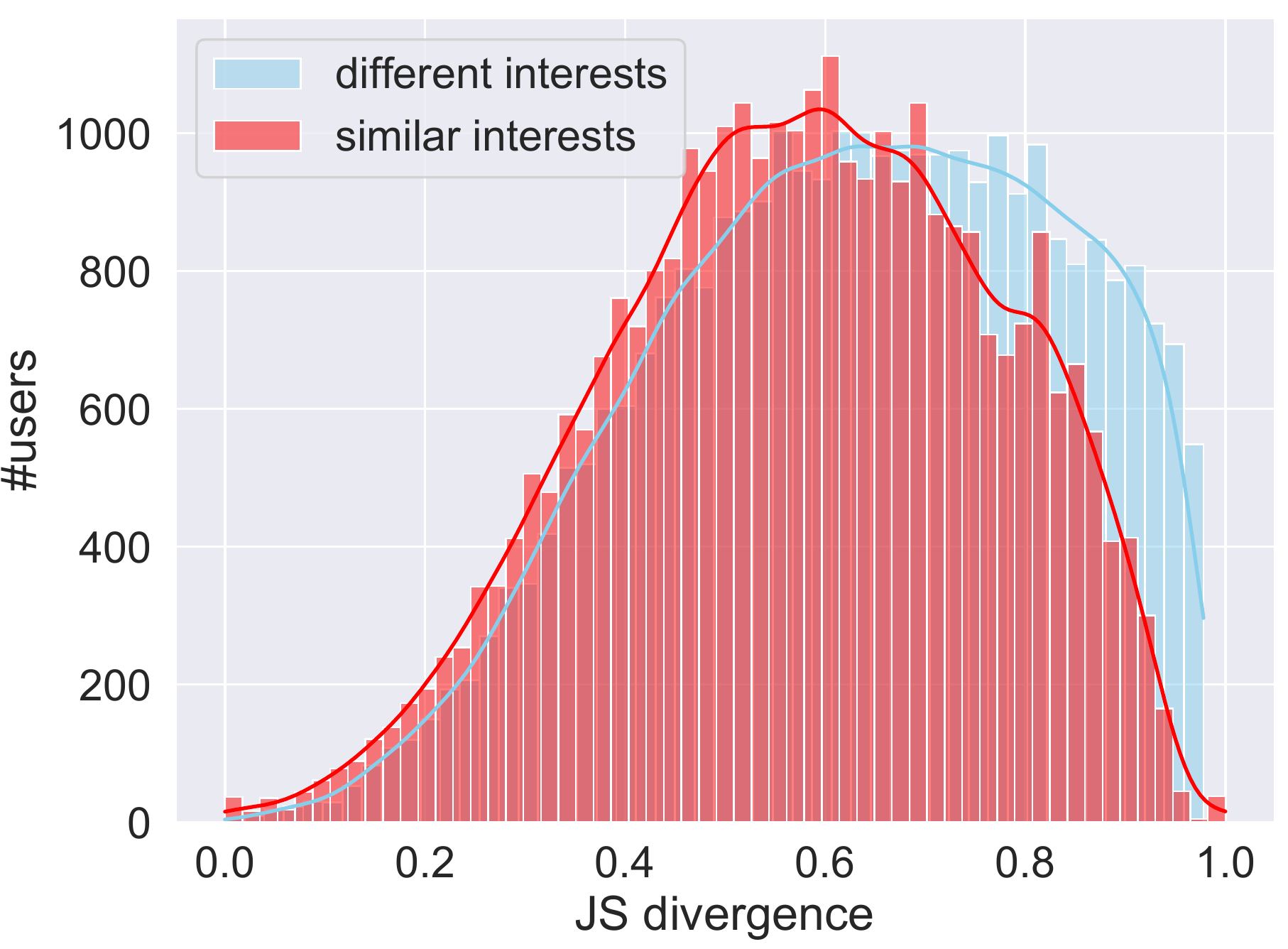}
    \vspace{-5px}
    \caption{Visualization of similarity between similar and dissimilar interests in S\&R behaviors for all users based on a histogram.
    We use the JS divergence of item category distributions to estimate similarities between S\&R behaviors.
    Similar interests are more similar than dissimilar interests with smaller values of JS divergence.
    }
\label{fig:js}
\end{figure}

The core operation of interest disentanglement is the separation of similar and dissimilar interests. 
To illustrate whether the interests are disentangled, we visualize and interpret the similarity between constructed sets of similar and dissimilar interests in~\autoref{fig:js}.
Behaviors  $\mcal{P}_s,\mcal{P}_r$ containing similar interests and $\mcal{N}_s, \mcal{N}_r$ containing dissimilar interests are  separated in Equation~\eqref{eq:separated1}~and~\eqref{eq:separated2}.
For each user, we calculated the Jensen–Shannon (JS) divergence~\cite{lin1991divergence} between sets with similar interests, \ie $D_{JS}(\mcal{P}_s\|\mcal{P}_r)$, which is colored in red in~\autoref{fig:js}.
The same calculation also was done for sets with dissimilar interests, \ie $D_{JS}(\mcal{N}_s\|\mcal{N}_r)$, which is colored in blue in~\autoref{fig:js}.
As discussed in~\autoref{sec:intro}, we can use the categories of items $S_i^u$ (interacted in recommendation history) and $S_c^u$ (clicked in search history) to estimate the similarities between two behaviors.
The calculation of JS divergence is based on the distributions of item categories corresponding to $\mcal{P}_s, \mcal{P}_r$ and $\mcal{N}_s,\mcal{N}_r$ for each user.
\autoref{fig:js} illustrates the results for all users.
We observed that similar interests tend to have smaller values of JS divergence than dissimilar interests, where red data has more counts smaller than 0.6 compared with blue data.
This phenomenon indicates $\mcal{P}_s$ and $\mcal{P}_r$ are more similar than $\mcal{N}_s$ and $\mcal{N}_r$, verifying the capability of SESRec to disentangle user interests.

\subsubsection{Analysis of $\gamma_s, \gamma_r$ in Interest Disentanglement.}

\begin{table}
\centering
    \caption{The analysis of the positive/negative selection thresholds $\gamma_r, \gamma_s$ in interest disentanglement, as defined in Equation~\eqref{eq:separated1}~and~\eqref{eq:separated2}.
    }
    \vspace{-10px}
{
\renewcommand{\arraystretch}{1.1}
\begin{tabular}{c ccc ccc}
\toprule
$\gamma_r, \gamma_s$
   & N@5 & N@10 & H@1 & H@5 &H@10 &MRR \\
    \midrule
    $1/16$   &  0.3429  & 0.3927 &   0.2125  &  0.4681   &  0.6224    &  0.3429      \\
    $1/8$    &  0.3449  &  0.3940   &  0.2148   &   0.4691  &  0.6211  &   0.3415       \\
    Median   &   0.3510   &   0.4035   &  0.2135     &   0.4828   &  
    \textbf{0.6453}    &   0.3462        \\
    Mean    &   \textbf{0.3541}   &   \textbf{0.4054}   &   \textbf{0.2173}    &   \textbf{0.4848}   &  0.6436    &  \textbf{0.3490}         \\
    \bottomrule
\end{tabular}
}%
    \label{tab:analysis of thresholds}
\end{table}

The separation of similar and dissimilar interests depends on the positive/negative behaviors selection, as defined in Equation~\eqref{eq:separated1}~and~\eqref{eq:separated2}.
We investigated the impacts of different positive/negative selection thresholds $\gamma_r, \gamma_s$.
In practice, we utilize an adaptive way, \ie setting thresholds as mean values ($\gamma_r = \frac{1}{T_r},\gamma_s=\frac{1}{T_r}$), to choose the positives and negatives.
To show how $\gamma_r, \gamma_s$ affect the performance of SESRec, we conducted experiments where $\gamma_r, \gamma_s$ are set as constants, \eg $\frac{1}{8}$ and $\frac{1}{16}$, and we also investigated another adaptive way that sets $\gamma_r, \gamma_s$ as the median values of given similarity scores.
\autoref{tab:analysis of thresholds} presents the results with different settings of $\gamma_r, \gamma_s$.
We can observe that constant settings ($\gamma_r, \gamma_s = \frac{1}{8}$ or $\frac{1}{16}$) yield inferior performances compared with the two adaptive strategies.
We postulate that the constant settings can not handle behaviors with different lengths consistently because longer behaviors lead to smaller mean values of scores normalized by $\mrm{softmax}$.
We also find that the adopted mean value strategy achieves better performance than the median value strategy in most cases.
The median value strategy separates behavior sequences into two parts of the same length.
However, the similar and dissimilar parts of users' behaviors do not satisfy this distribution in most cases.
That is why the adopted mean value strategy achieves the best performance.

\subsubsection{Effect of Query-item Alignment.}

\begin{figure}
    \centering
        \includegraphics[width=0.8\linewidth]{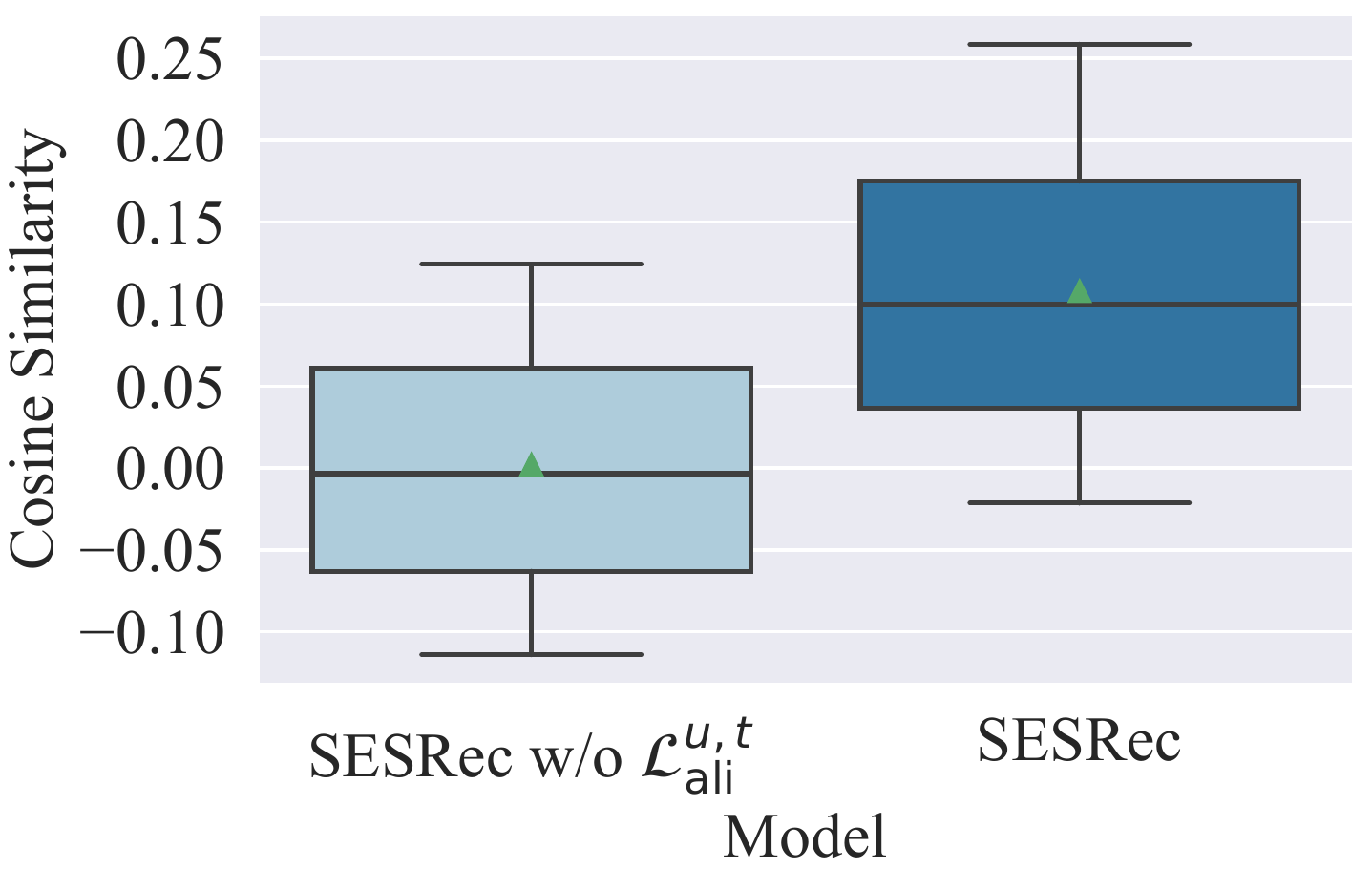}
    \vspace{-10px}
    \caption{Distribution of cosine similarity between representations of queries and corresponding items based on box plots. Rectangles denote mean values.
    With query-item alignment loss $\mcal{L}^{u,t}_{\mrm{ali}}$, embeddings of query-item pairs are more similar with higher cosine values.
    }
\label{fig:cos}
\vspace{-5px}
\end{figure}

We conducted experiments to explore how the query-item alignment facilitates representation learning and whether SESRec understands the similarity of queries and items.
Toward this end, we tested the relevance between queries and their clicked items.
For search behaviors, we split queries and their clicked items into pairs, where each pair consists of a query and its corresponding item.
And we obtained their embeddings learned by SESRec and a variation of SESRec, which removes the query-item alignment loss $\mathcal{L}^{u,t}_{\mrm{ali}}$.
We calculated the cosine similarity of each query-item pair based on the embeddings and plotted the distribution of similarity scores in~\autoref{fig:cos}.
From the results, we can observe that embeddings learned by SESRec have smaller similarity scores than those learned without $\mathcal{L}^{u,t}_{\mrm{ali}}$.
These results indicate that the query-item alignment module ensures that queries are closed to their corresponding items in correlation. 

\subsubsection{Impact of Hyper-parameters.}

\begin{figure}
    \centering
    \begin{subfigure}{0.486\linewidth}
        \includegraphics[width=\textwidth]{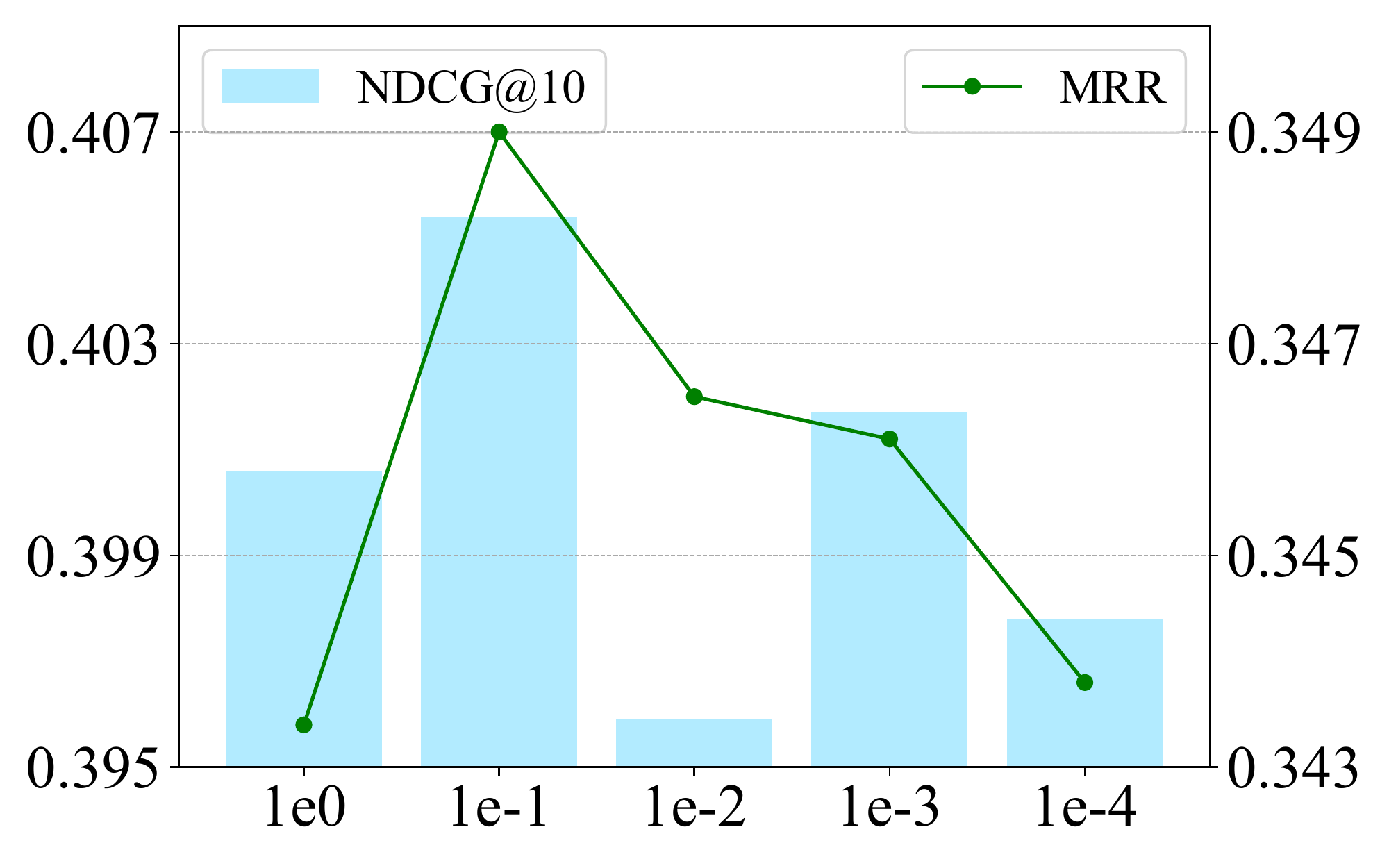}
        \subcaption{Performance of different $\alpha$.}
    \end{subfigure}
    \begin{subfigure}{0.486\linewidth}
        \includegraphics[width=\textwidth]{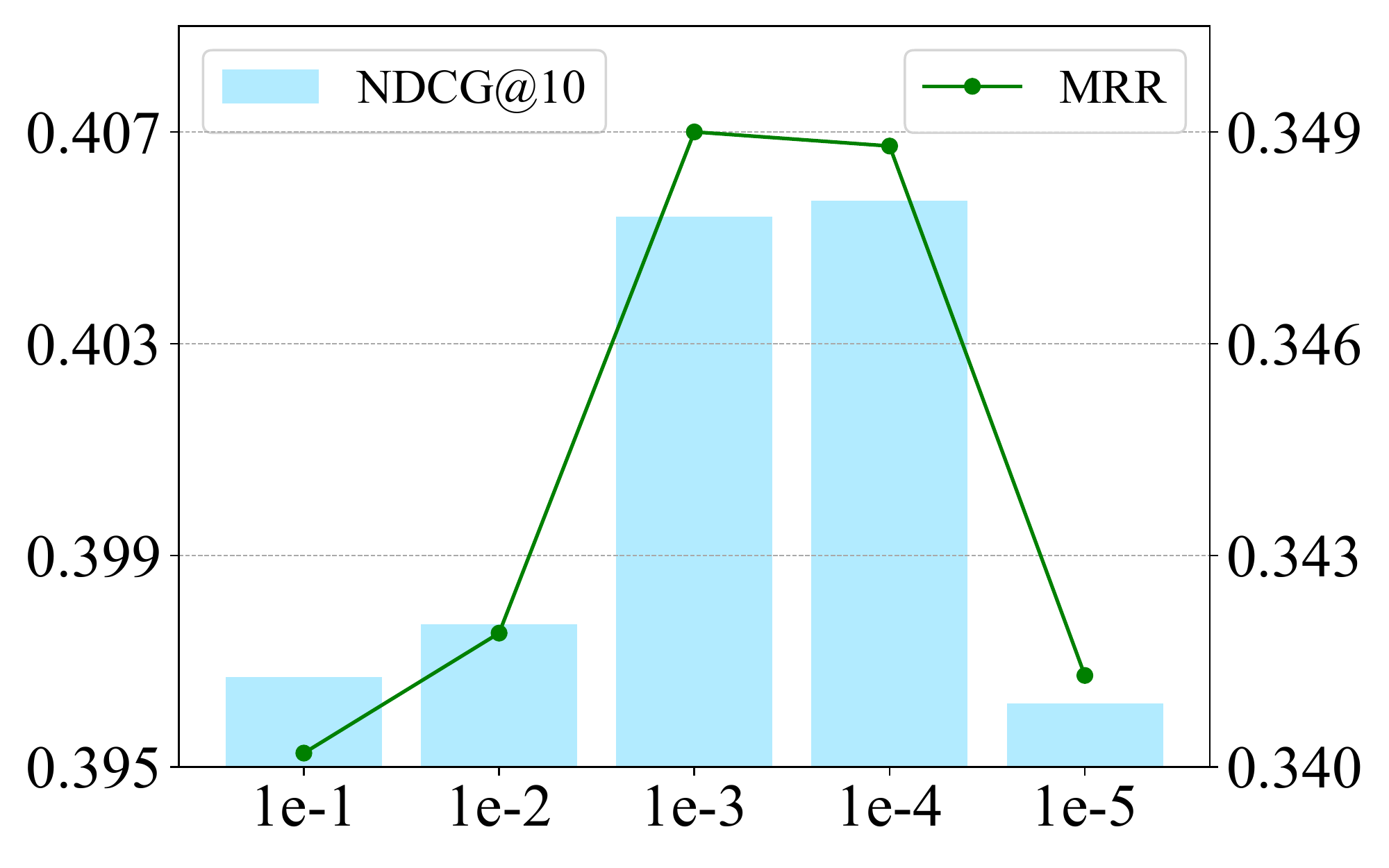}
        \subcaption{Performance of different $\beta$.}
    \end{subfigure}
    \vspace{-5px}
    \caption{Effects of hyper-parameters $\alpha$ and $\beta$ in terms of NDCG@10 and MRR.}
\label{fig:hy-para}
\vspace{-5px}
\end{figure}

Since we design two additional tasks, hyper-parameters $\alpha$ and $\beta$ are introduced to balance the objectives in the final loss function, as defined in Equation~\eqref{eq: loss func}.
To investigate the impacts of these hyper-parameters, we conducted experiments with varying $\alpha$ and $\beta$ respectively.
When varying one parameter, the other is set as a constant, where 1e-1 for $\alpha$ and 1e-3 for $\beta$.
From the results in~\autoref{fig:hy-para}, we found that the performance peaks when $\alpha$ is 1e-1 and $\beta$ is 1e-3.
With a further increase of hyper-parameters, the recommendation performances become worse.
We attribute it to the fact that the recommendation prediction task becomes less important with larger $\alpha$ and $\beta$, which verifies the necessity of hyper-parameters to balance different tasks in the multi-task learning schema.

%% file: conclusion.tex
\section{Conclusion}
In this paper, we propose to learn disentangled search representation for recommendation with a search-enhanced framework, namely SESRec.
SESRec exploits the query-item interactions to help the recommendation model to learn better representations of queries and items.
With the help of self-supervision, SESRec disentangles the similar and dissimilar representations between users' search and recommendation behaviors to capture users' interests from multiple aspects.
Besides, SESRec provides an end-to-end multi-task learning framework for estimating the parameters.
Extensive experiments on industrial and public datasets demonstrate that SESRec consistently outperforms state-of-the-art baselines.
Moreover, we empirically validate that SESRec successfully learn the disentangled representations of user interests.